\documentclass[useAMS,usenatbib,twocolumns]{mn2e}
\usepackage{graphicx}
\usepackage{multirow}
\usepackage{lipsum}
\usepackage{blindtext}
\usepackage{amsmath}
\usepackage{amssymb}
\usepackage[usenames, dvipsnames]{color}

\newcommand{\OII}{$\left[\mathrm{O\,\textrm{\textsc{ii}}}\right]$\xspace}
\newcommand{\OIIII}{$\left[\mathrm{O\,\textrm{\textsc{ii}}}\right]\,(\lambda\lambda3726,3729)$\xspace}
\newcommand{\NII}{$\left[\mathrm{N\,\textrm{\textsc{ii}}}\right]$\xspace}
\definecolor{mygray}{gray}{0.5}

\title[\OII ELG clustering at $z\sim0.1$]{Galaxy clustering dependence on the [O\textsc{II}] emission line luminosity in the local Universe }
\author[Favole et al. 2017]
{\parbox[t]{\textwidth}{\vspace{-0.7cm}Ginevra Favole$^{1}$\thanks{E-mail: gfavole@sciops.esa.int}, Sergio A. Rodr\'iguez-Torres$^{2,3,4}$\thanks {Campus de Excelencia Internacional UAM/CSIC Scholar}, Johan Comparat$^{2,3,4}$\thanks {Severo Ochoa IFT Fellow}, Francisco Prada$^{2,3,5}$, Hong Guo$^{6}$, Anatoly Klypin$^{7}$, Antonio D. Montero-Dorta$^8$}
\vspace*{12pt}\\ 
$^1$European Space Astronomy Center (ESAC), 28692 Villanueva de la Ca\~nada, Madrid, Spain\\
$^2$Instituto de F\'{i}sica Te\'{o}rica (IFT) UAM/CSIC, Universidad Aut\'{o}noma de Madrid, Cantoblanco, E-28049 Madrid, Spain\\
$^3$Campus of International Excellence UAM/CSIC, Cantoblanco, E-28049 Madrid, Spain\\
$^4$Departamento de F\'isica Te\'orica M-8, Universidad Aut\'onoma de Madrid, Cantoblanco, 28049 Madrid, Spain\\
$^5$ Instituto de Astrof\'isica de Andaluc\'ia (CSIC), Granada, E-18008, Spain \\
$^6$Key Laboratory for Research in Galaxies and Cosmology, Shanghai Astronomical Observatory, Shanghai 200030, China\\
$^7$Astronomy Department, New Mexico State University, MSC 4500, P.O. Box 30001, Las Cruces, NM, 880003-8001, USA\\
$^8$Department of Physics and Astronomy, University of Utah, UT 84112, USA
\vspace{-1cm}
}
\date{  }
\begin{document}
\pagerange{\pageref{firstpage}--\pageref{lastpage}} \pubyear{2016}
\maketitle

\begin{abstract}

\noindent We study the galaxy clustering dependence on the \OII emission line luminosity in the SDSS DR7 Main galaxy sample at mean redshift $z\sim0.1$. We select volume-limited samples of galaxies with different \OII luminosity thresholds and measure their projected, monopole and quadrupole two-point correlation functions. We model these observations using the 1$h^{-1}$Gpc MultiDark Planck cosmological simulation and generate light-cones with the SUrvey GenerAtoR algorithm. To interpret our results, we adopt a modified (Sub)Halo Abundance Matching scheme, accounting for the stellar mass incompleteness of the emission line galaxies. The satellite fraction constitutes an extra parameter in this model and allows to optimize the clustering fit on both small and intermediate scales (i.e. $r_p\lesssim 30\,h^{-1}$Mpc), with no need of any velocity bias correction. We find that, in the local Universe, the \OII luminosity correlates with all the clustering statistics explored and with the galaxy bias. This latter quantity correlates more strongly with the SDSS $r$-band magnitude than \OII luminosity. In conclusion, we propose a straightforward method to produce reliable clustering models, entirely built on the simulation products, which provides robust predictions of the typical ELG host halo masses and satellite fraction values. The SDSS galaxy data, MultiDark mock catalogues and clustering results are made publicly available.
\end{abstract}


\section{Introduction}
\label{sec:intro}

In the last  decade, the Sloan Digital Sky Survey \citep[SDSS;][]{York2000, Gunn2006, Smee2013} first, and then the SDSS-III/Baryon Oscillation Spectroscopic Survey \citep[BOSS;][]{Eisenstein2011, Dawson2013} have mainly observed luminous red galaxies \citep[LRGs;][]{Eisenstein2001} up to redshift $z\sim0.7$ to trace the baryon acoustic oscillation feature \citep[BAO;][]{Eisenstein2005} in their clustering signal and use it as \textquoteleft \textquoteleft standard ruler" for cosmological distances. 
New-generation large-volume spectroscopic surveys, both ground-based as eBOSS \citep[]{2016AJ....151...44D}, DESI \citep[]{2015AAS...22533607S}, 4MOST \citep[]{2012EPJWC..1909004D}, Subaru-PFS \cite[]{2015JATIS...1c5001S} and space-based as Euclid \citep[]{2011arXiv1110.3193L, 2015arXiv150502165S}, have all been designed to probe larger volumes by looking back in time out to $z\sim2$ and  to target high-redshift star-forming galaxies with strong nebular emission lines (ELGs) as BAO tracers. Emission line galaxies that have the peak of star formation around $z\sim2$ are usually faint targets and their spectral features are shifted to the near-infrared (NIR) region. This makes them hard to detect using the optical ground-based facilities currently available \citep[]{2014ApJ...785..153M}. For this reason, only few studies \citep[e.g.,][]{2006ApJ...644..813E,2010ApJ...719.1168E,2009ApJ...701...52H,2011ApJ...732...59R,2013ApJ...772..141B, 2013ApJ...763..145D} of the ELG spectral properties have been conducted in the optical regime to date. Star-forming [OII] and H$\alpha$ emitters have been explored at NIR wavelengths by the slitless grim WFC3 Infrared Spectroscopic Parallel Survey \citep[WISP;][]{2010ApJ...723..104A} on the Hubble Space Telescope. This mission was designed to observe emission line galaxies with ongoing star formation out to $z\sim2$, and revealed several differences in comparison with the low-redshift scenario \citep[]{2008ApJ...678..758L, 2014ApJ...781...21N}. Star-forming galaxies at high redshift show higher [OIII]/H$\beta$ ratios, probably due to harder conditions in the interstellar medium \citep[see e.g.][]{2005ApJ...635.1006S, 2008MNRAS.385..769B,2014ApJ...787..120S,2013ApJ...774..100K}; they also have higher velocity dispersion \citep[]{2001ApJ...554..981P,2008ApJ...687...59G,2009ApJ...697.2057L}, which could be motivated by star formation taking place in denser environments in the early Universe \citep[]{2014ApJ...785..153M}. Despite its numerous successes, WISP lacks of spectral resolution to resolve H$\alpha$ from [NII] $\lambda=6548,\,6583$\,\AA, or to detect broad emission lines caused by active galactic nuclei.
The Euclid slitless grim spectroscopic program, with a similar design to WISP, will be able to measure the \NII/H$\alpha$ flux ratio with small enough errors to reliably distinguish narrow-line AGN and star-forming galaxies down to H$\alpha$ fluxes of $\sim1.5\times10^{-15}\,\rm{erg\,cm^{-2}\,s^{-1}}$, over the redshift range $0.7<z<2$ \citep[]{2011arXiv1110.3193L}.
Complementing the new generation of ground-based infrared spectrometers of eBOSS and DESI with the Euclid space-based facility will help to constrain the physical properties of these emission line galaxies at high redshift, and to understand better their formation and evolution processes.

\noindent In preparation to these near-future experiments, we discuss here how to measure and correctly model the ELG distribution within their host haloes, their clustering properties as a function of the emission line luminosity and its evolution with redshift, using the data currently available. Specifically, we focus our analysis on the SDSS DR7 Main galaxy sample \citep[]{2002AJ....124.1810S, 2009ApJS..182..543A} at mean redshift $z\sim0.1$.
We select this sample from the New York-Value Added Galaxy catalogue \citep[NYU-VAGC][]{2005AJ....129.2562B} and assign \OII luminosities by performing a spectroscopic matching to the MPA-JHU DR7 (at \url{http://www.mpa-garching.mpg.de/SDSS/DR7/})  release of spectrum measurements. In the merged galaxy population, we select volume-limited samples in different \OII luminosity thresholds, where we measure the projected, monopole and quadrupole two-point correlation functions (2PCFs) and model the results in terms of the typical host halo masses and satellite fraction. We adopt a (Sub)Halo Abundance matching \citep[SHAM;][]{Klypin2010,Trujillo2011} scheme, modified \citep[]{2017MNRAS.468..728R} to account for the ELG stellar mass incompleteness. The method applied is straightforward since entirely built on the simulation products, with no need of introducing any velocity bias \citep[]{2015MNRAS.446..578G}, and provides reliable predictions of the ELG host halo masses and satellite fraction values, as a function of the \OII luminosity. 
Our galaxy data and mock catalogues, together with the clustering results, are publicly available both on the \textsc{Skies and Universes} database at \url{http://projects.ift.uam-csic.es/skies-universes/SUwebsite/indexSDSS_OII_mock.html}, and as \textsc{MNRAS} online material.

\noindent The paper is organized as follows: in Section \ref{sec:dataall} we describe the SDSS data set used and how we assign \OII luminosities. In Section \ref{sec:measurements} we introduce the tools needed to perform our clustering measurements. In Section \ref{sec:models} we present the simulation and the ELG clustering model. In Section \ref{sec:results} we present our ELG clustering results and the correlation between \OII luminosity and galaxy bias. We discuss our conclusions and the future plans in Section \ref{sec:disc}. 
In what follows, we adopt the \citet[]{Planck2014} cosmology: $\Omega_m=0.3071$, $\Omega_{\Lambda}=0.6929$, $h=0.6777$, $n=0.96$, $\sigma_8=0.8228$.

\section{Data}
\label{sec:dataall}

We select the SDSS DR7 Main galaxy sample \citep[]{2002AJ....124.1810S} from the NYU-VAGC\footnote{\url{http://cosmo.nyu.edu/blanton/vagc/}} \citep[]{2005AJ....129.2562B} and assign \OII emission line fluxes by performing a spectroscopic matching to the MPA-JHU\footnote{\url{http://www.mpa-garching.mpg.de/SDSS/DR7/}} DR7 release of spectrum measurements. We consider only MPA-JHU galaxies with good spectra, i.e. with \texttt{ZWARNING=0}.
In what follows, the merged galaxy catalogue will be called ``MPA-NYU SDSS Main'' catalog. Notice that all the galaxies in this sample show \OII emission lines, meaning that we do not include any elliptical or quiescent galaxy which could be central for some of the ELGs considered.  We compute the \OII fluxes as $F=F_c\times |EQW|$,  
where $F_c$ is the line flux continuum and $EQW$ is equivalent width of the MPA-JHU lines. In the case of line doublets emitting two different wavelengths as \OII, the flux considered is the cumulative flux of both lines. Following \citet{2003ApJ...599..971H}, we correct these fluxes for extinction using \citet{Schlegel1998} $E(B-V)$ dust maps  and \citet{Calzetti2000} law to obtain:
\begin{equation}
F_{ext}^{corr}\,[\rm{erg\,\,s^{-1}cm^{-2}}]=F\times 10^{-0.4A_{\lambda}}=F\times 10^{-0.4E(B-V)k(\lambda_{obs})}.
\label{eq:Fobs}
\end{equation}
In the equation above, the quantity $k(\lambda_{obs})$ is the reddening curve \citep[]{Calzetti2000}, $\lambda_{obs}=\lambda_{em}(1+z)$ is the observed wavelength and $\lambda_{em}$ is the emitted one. 
The observed ELG luminosity is then recovered from the flux as \cite[][]{2003ApJ...599..971H}
\begin{equation}
L_{obs}\,[\rm{erg\,\,s^{-1}}]=4\pi D_L^2 10^{-0.4(m_p-m_{fib})}F_{ext}^{corr},
\label{eq:flux}
\end{equation}
where $D_L$ is the luminosity distance depending on redshift and cosmology. In Eq.\,\ref{eq:flux}, the exponent is the aperture correction accounting that only the portion of the flux ``through the fiber'' will be detected by the SDSS spectrograph -- fibers in SDSS have an aperture of 3'' \citep[]{2002AJ....124.1810S}.
This correction implicitly assumes that the emission measured through the fiber is characteristic of the whole galaxy and that the star formation is uniformly distributed over the galaxy.
The term $m_p$ is the petrosian magnitude in the desired band-pass filter representing the total galaxy flux and $m_{fib}$ is the fiber magnitude derived from a photometric measurement of the magnitude in an aperture the size of the fiber and corrected for seeing effects. In the SDSS $ugriz$ \citep[]{Gunn1998, 1996AJ....111.1748F} optical photometric system, the \OIIII doublet lies in the $u$-band.

\begin{figure*}
\centering 
  \includegraphics[width=0.78\linewidth]{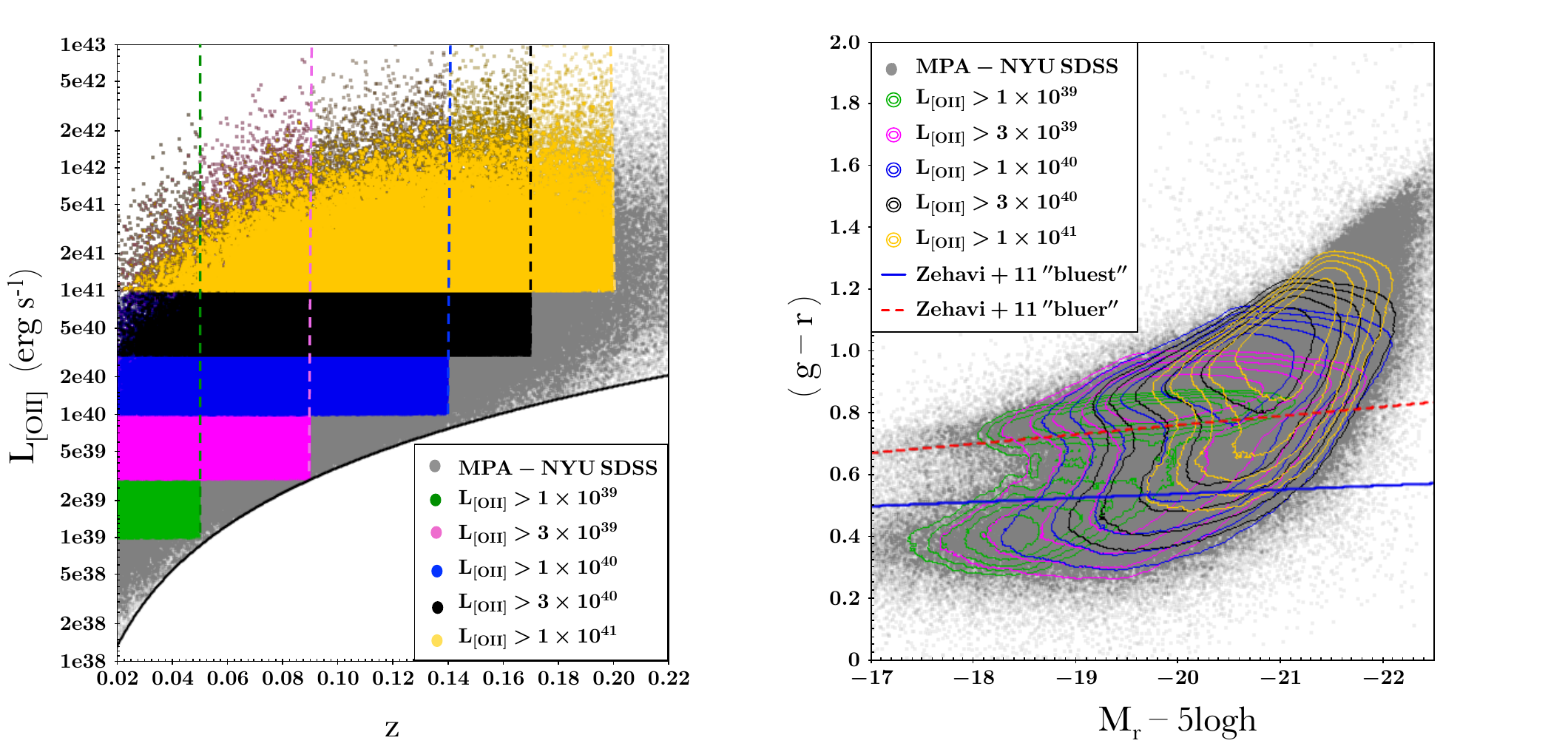}
\caption{\textit{Left:} SDSS \OII emission line luminosity (grey dots) and volume-limited samples (colored squares) for the NYU-MPA Main galaxies. We impose a conservative minimum flux limit (black line) of $F_{[\rm{OII}]}=3\times10^{-16}$erg\,cm$^{-2}$\,s$^{-1}$ to exclude objects with too short exposure time. \textit{Right:} SDSS \OII ELG color-magnitude diagram. Our result is fully compatible with the ``bluer'' (dashed horizontal line) and ``bluest'' (solid) SDSS sub-samples defined by \citet[]{2011ApJ...736...59Z}.}
  \label{fig:samplesO2}
  \end{figure*}

\noindent The observed \OII emission line luminosity of the NYU-MPA SDSS Main galaxies in the redshift range $0.02<z<0.22$ is displayed in the left panel of Figure \ref{fig:samplesO2}. In the right panel we show the $(g-r)$ \OII ELG color as a function of the SDSS $r$-band absolute magnitude, which is in agreement with the ``bluer'' and ``bluest'' sub-samples in \citet[]{2011ApJ...736...59Z}. Since the SDSS DR7 spectra are combined from three or more individual exposures of 15 minutes each (see \url{http://classic.sdss.org/dr7/products/spectra/}), corresponding to typical \OII fluxes of $\sim10^{-16}\rm{\,erg\,cm^{-2} s^{-1}}$ \citep[]{2015A&A...575A..40C}, we impose a conservative limit rejecting all galaxies with \OII flux lower than $3\times10^{-16}\rm{\,erg\,cm^{-2} s^{-1}}$ (black line in Figure \ref{fig:samplesO2}) and remain with a sample of about 427,000 \OII ELGs. 
To investigate the ELG clustering dependence on the \OII luminosity, we select several volume-limited samples in $L_{\rm{[OII]}}$ thresholds and show them as colored squares in Figure \ref{fig:samplesO2}. The specific cuts to obtain them are reported in Table \ref{tab:vlsO2}.

\begin{table*}
  \centering
  \begin{tabular}{c c c c c c c c c c }
    \hline
    $z_{max}$&L$_{\rm{[O_{II}]}}^{min}$&$N_{gal}$&$\bar{n}_g$&Vol&$f_{\rm{sat}}$&$V_{\rm{peak}}\pm\sigma_V$&typical $V_{\rm{peak}}$&typical M$_{h}$ &$\chi^2$\\
            &[erg s$^{-1}$]&&[$10^{-3}h^{3}$Mpc$^{-3}$]&[$10^6h^{-3}$Mpc$^3$]&[\%]& [km\,s$^{-1}$]&[km\,s$^{-1}$]&[$h^{-1}$M$_{\odot}$]&\\
    \hline
    0.05&$1\times10^{39}$&57595  &25.60&$\,\,$2.25     &33.4$\pm$0.1&275$\pm$145&127$\pm$58&(3.17$\pm$0.19)$\times10^{11}$&1.82\\
  0.09&$3\times10^{39}$&174360&12.92&13.50  &27.9$\pm$0.4&285$\pm$130&177$\pm$53&(6.64$\pm$0.41)$\times10^{11}$&2.37\\
  0.14&$1\times10^{40}$&244700&4.95  &49.39  &22.5$\pm$0.7&310$\pm$107&201$\pm$86&(1.54$\pm$0.09)$\times10^{12}$&3.62\\
  0.17&$3\times10^{40}$&184622&2.16  &85.59  &19.4$\pm$0.4&284$\pm$131&$\,\,\,$283$\pm$117&(2.92$\pm$0.18)$\times10^{12}$&2.17\\
  0.20&$1\times10^{41}$&89814  &0.65  &137.91&18.0$\pm$0.5&303$\pm$140&$\,\,\,$341$\pm$140&(5.49$\pm$0.34)$\times10^{12}$&5.08\\
    \hline
  \end{tabular}
   \caption{Redshift and \OII luminosity cuts defining the MPA-NYU SDSS Main volume-limited samples. For each sample we report the number of galaxies ($N_{gal}$) contained, its number density ($n_g$), and its comoving volume (Vol). We impose a minimum redshift of $z=0.02$ and a minimum \OII line flux of $3\times10^{-16} \rm{erg\,cm^{-2}\,s^{-1}}$ to each one of the samples (see text for details). We also show our predictions (see Section \ref{sec:results}) of the satellite fraction (in units of percent), the best-fit $V_{\rm{peak}}\pm\sigma_V$ and the $\chi^2$ values of our ELG clustering model. For the $w_p(r_p)$ fits we use 11 dof. In addition, we display the ``typical''  halo velocity and mass values derived from the resulting SDSS \OII ELG mock catalogues. These quantities are the values at the peak of the final  $V_{\rm{peak}}$ and M$_h$ distributions, which characterize the haloes that \OII ELGs most probably  occupy in the local Universe.}
 \label{tab:vlsO2}
\end{table*}

\section{Clustering measurements}
\label{sec:measurements}

\noindent We measure the projected, monopole and quadrupole two-point correlation functions  of the NYU-MPA SDSS Main volume-limited samples using the Landy-Szalay estimator \citep[]{1993ApJ...412...64L}, for which we build suitable randoms including the angular and radial footprint of the data samples. We take into account the variation of the completeness across the sky by downsampling the NYU-VAGC 
random catalogue with equal surface density in a random fashion using the completeness 
as a probability function. We then shuffle the (RA, DEC) random coordinates by sorting them with respect to a random flag, and assign the observed redshifts \citep[]{2014MNRAS.441...24A}.
To correct for angular incompleteness, we weight each object in the real and random catalogues by $w_{\rm{ang}}$, 
which is defined as the inverse of the SDSS sector completeness. Since the minimum angular distance allowed between the SDSS optical fiber is 55" (i.e. $r_p\sim\,0.13\,h^{-1}$Mpc at $\bar{z}=0.1$), we limit the measurements to scales larger than that and upweight by one \cite{Ross2012} the fiber collision \citep[]{Zehavi2002, 2006ApJ...644...54M} weight, $w_{fc}$, of the nearest neighbor of the collided galaxy. Because we use volume-limited samples, we do not need to apply any radial weight. We combine the corrections above into a total weight \citep[]{Sanchez2012} of $w_{\rm{tot}}=w_{\rm{fc}}\,w_{\rm{ang}}$.
To estimate the errors on our clustering measurements we perform 200 jackknife \citep[e.g.,][]{Turkey1958, Norberg2011, Ross2012, Anderson2012} re-samplings. 

\section{Simulation and modeling}
\label{sec:models}
 \begin{figure*}
\centering 
  \includegraphics[width=0.35\linewidth]{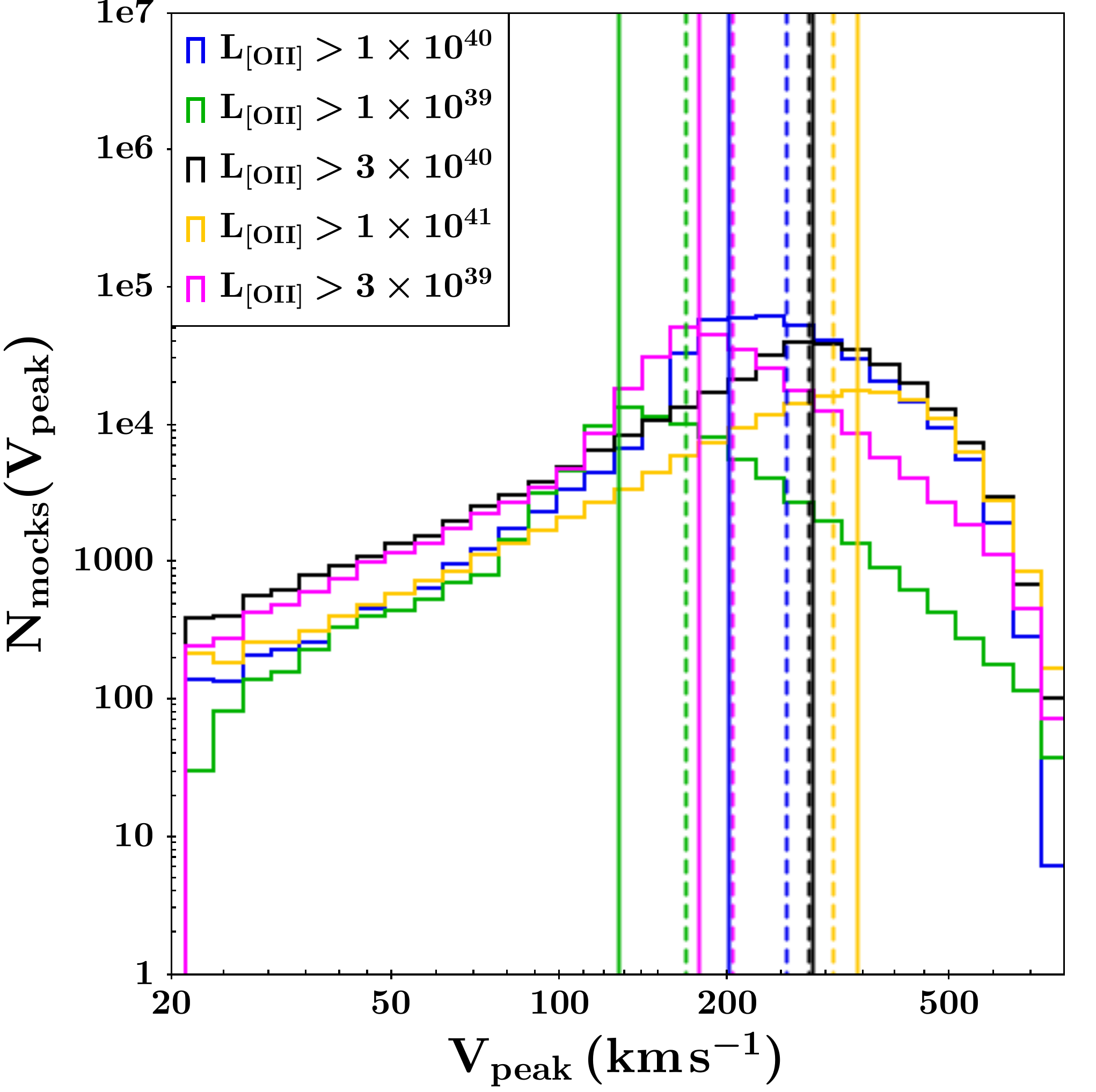}\hspace{0.5cm}
  \includegraphics[width=0.35\linewidth]{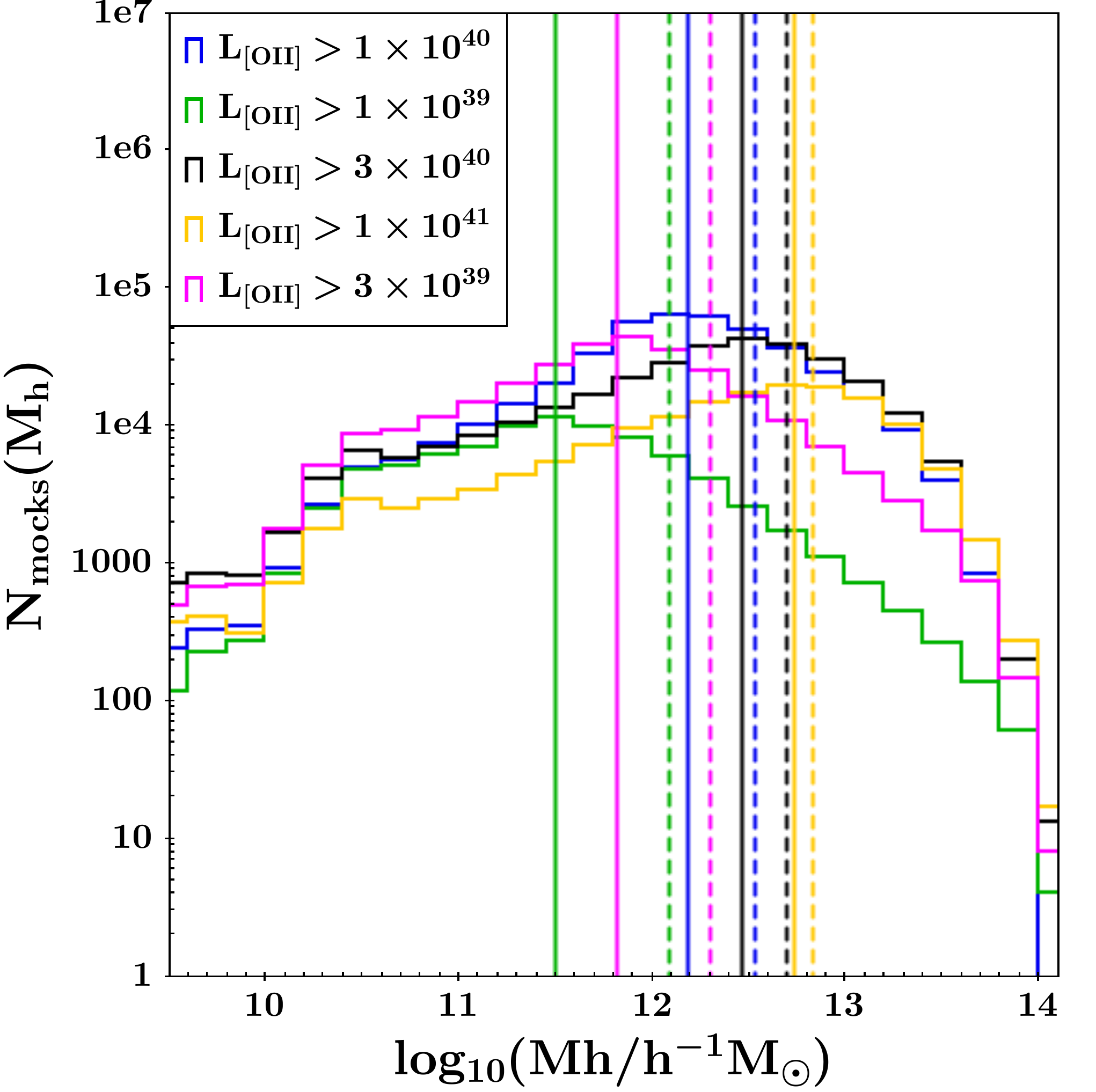}
\caption{$V_{\rm{peak}}$ (left) and halo mass (right) \OII ELG mock distributions. The dashed vertical lines are the mean values, while the solid lines are the most probable values or ``typical'' values for haloes hosting \OII ELGs in the local Universe.}
  \label{fig:distrib}
  \end{figure*}

\noindent We model the ELG clustering measurements using the MultiDark\footnote{\url{https://www.cosmosim.org}} \citep[MDPL;][]{2016MNRAS.457.4340K} N-body cosmological simulation with Planck cosmology \citep[]{Planck2014}. 
The simulation box is $1\,h^{-1}$Gpc on a side and contains $3840^3$ particles with a mass resolution of $1.51\times10^9\,h^{-1}$M$_{\odot}$. It represents the best compromise between resolution and volume available to date.  
We apply the \textsc{SUrvey GenerAtoR} \citep[\textsc{SUGAR};][]{Rodriguez-Torres2016} algorithm to the MultiDark ROCKSTAR \citep[]{2013ApJ...762..109B} snapshots in the redshift range $0.02<z<0.22$ to produce suitable light-cones with about twice the area of the data (i.e., ${\rm{A_{LC}}}\sim12,000\,\deg^2$) to reduce the effect of cosmic variance. The advantage of using a light-cone instead of a single simulation snapshot at the mean redshift of the sample is that it includes the redshift evolution. In addition, it accounts for those volume effects (cosmic variance and galaxy number density fluctuations) that are observed in the data and a single simulation snapshot cannot capture. The disadvantage, however, is that the light-cone has a reduced volume compared to the original MDPL simulation box. To build it, we place its vertex at the center of one of the sides of the box and then we re-map this latter to determine the maximum light-cone aperture at $z=0.22$. From the $1\,h^{-1}$Gpc MultiDark simulation, we derive a cone with a volume of  $\sim0.02\,h^{-3}$Gpc$^3$, i.e. only 1/50 of the original box. In alternative, we could generate a light-cone with different apertures for each one of the redshift bins considered and obtain the full sky for the smallest $z$ interval. We could multiply by three the area of the current light-cone ($\sim$12,000 square degrees, i.e. 1/3 of the full sky) to have the full sky, but the gain in volume compared to the original $1\,h^{-1}$Gpc box would not be significant for most of the redshift slices considered. 

\noindent To  populate the MDPL light-cones with the galaxies of the SDSS MPA-NYU volume-limited samples, we adopt the modified (Sub)Halo Abundance Matching \citep[SHAM;][]{2011ApJ...740..102K,Trujillo2011} prescription proposed by \citet[]{2017MNRAS.468..728R} which accounts for the ELG stellar mass incompleteness, similarly to the method presented in \citet[]{Favole2016}. 
This SHAM assignment is performed by drawing mocks using two separate probability distribution functions (PDFs), one for central and one for satellite galaxies, both based on a Gaussian realization $\mathcal{G}_{s/c}(V_{\rm{peak}},z)$ with mean $V_{\rm{peak}}$ and standard deviation $\sigma_V$. Here $V_{\rm{peak}}$ is the halo maximum circular velocity over its entire history. The Gaussian realizations are normalized to reach the observed ELG number density and using the satellite fraction as a free parameter in the following way:
\begin{equation}
\begin{aligned}
    \int\mathcal{G}_s(V_{\rm{peak}},z)\,dV_{\rm{peak}}=N_{\rm{tot}}(z)f_{\rm{sat}}\\
 \int\mathcal{G}_c(V_{\rm{peak}},z)\,dV_{\rm{peak}}=N_{\rm{tot}}(z)(1-f_{\rm{sat}}).
\label{eq:normaliz}
\end{aligned}
\end{equation}
In the equation above, $N_{\rm{tot}}(z)$ is the total number of galaxies per redshift bin and is determined by their observed $n(z)$.
In practice, to build the PDFs (i) we sort the MDPL haloes, (ii) we separately compute the central and satellite halo velocity functions, (iii) we bin the haloes in $V_{\rm{peak}}$, (iv) we compute the probability of selecting a satellite/central mock as 
\begin{equation}
P_{\rm{s/c}}(V_{\rm{peak}})=\frac{N_{\rm{s/c}}^{\rm{Gauss}}}{N_{\rm{s/c}}^{\rm{tot}}},
\label{eq:pdfs}
\end{equation}
where $N_{\rm{s/c}}^{\rm{Gauss}}$ is the number of satellite/central mocks obtained by applying the Gaussian selection above, and $N_{\rm{s/c}}^{\rm{tot}}$ is the number of satellite/central haloes of the simulation in the redshift range considered.
For further details on the selection method see \citet[]{2017MNRAS.468..728R}.

\begin{figure*}
\centering
    \includegraphics[width=0.43\linewidth]{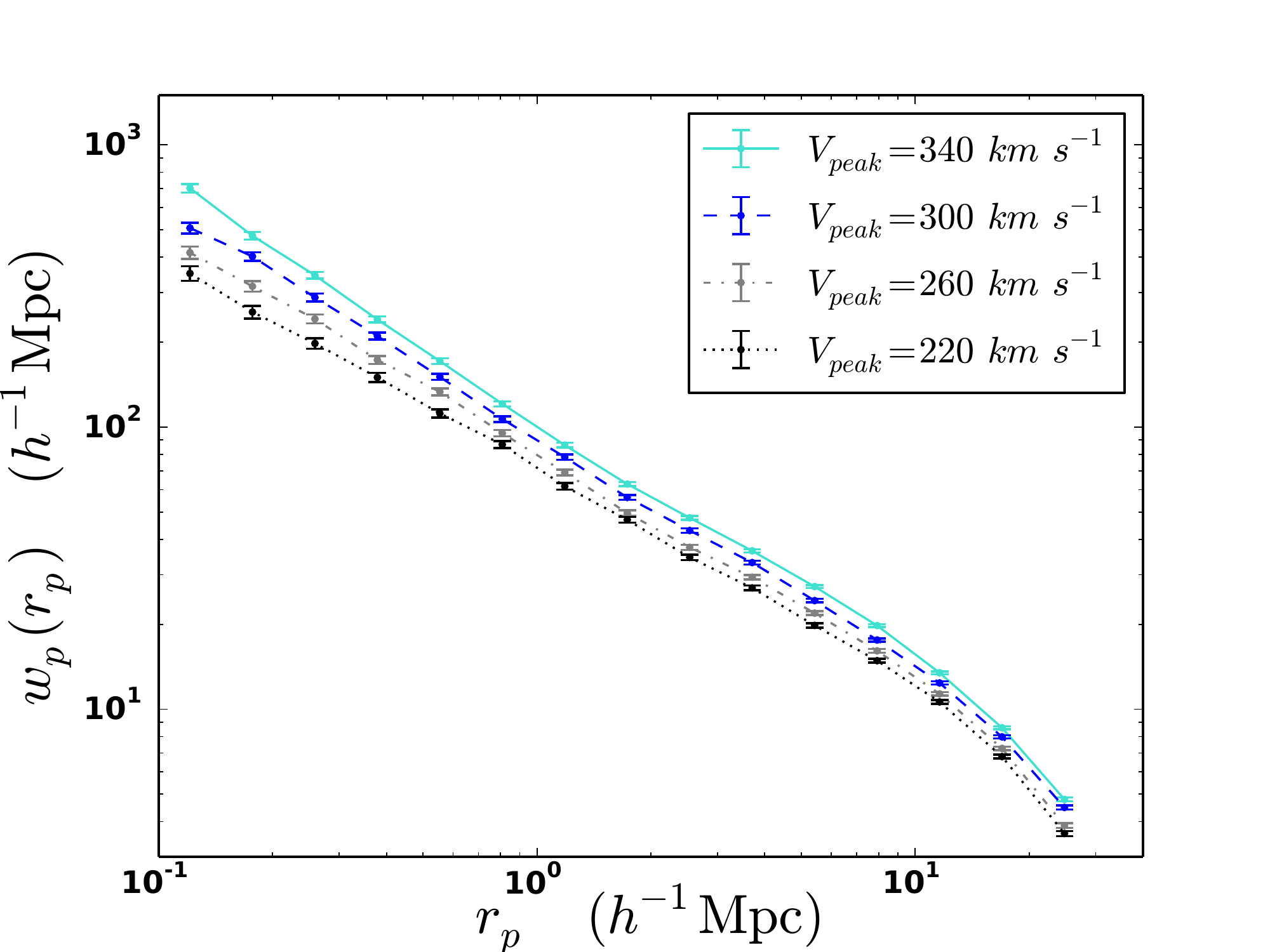}\quad
      \includegraphics[width=0.43\linewidth]{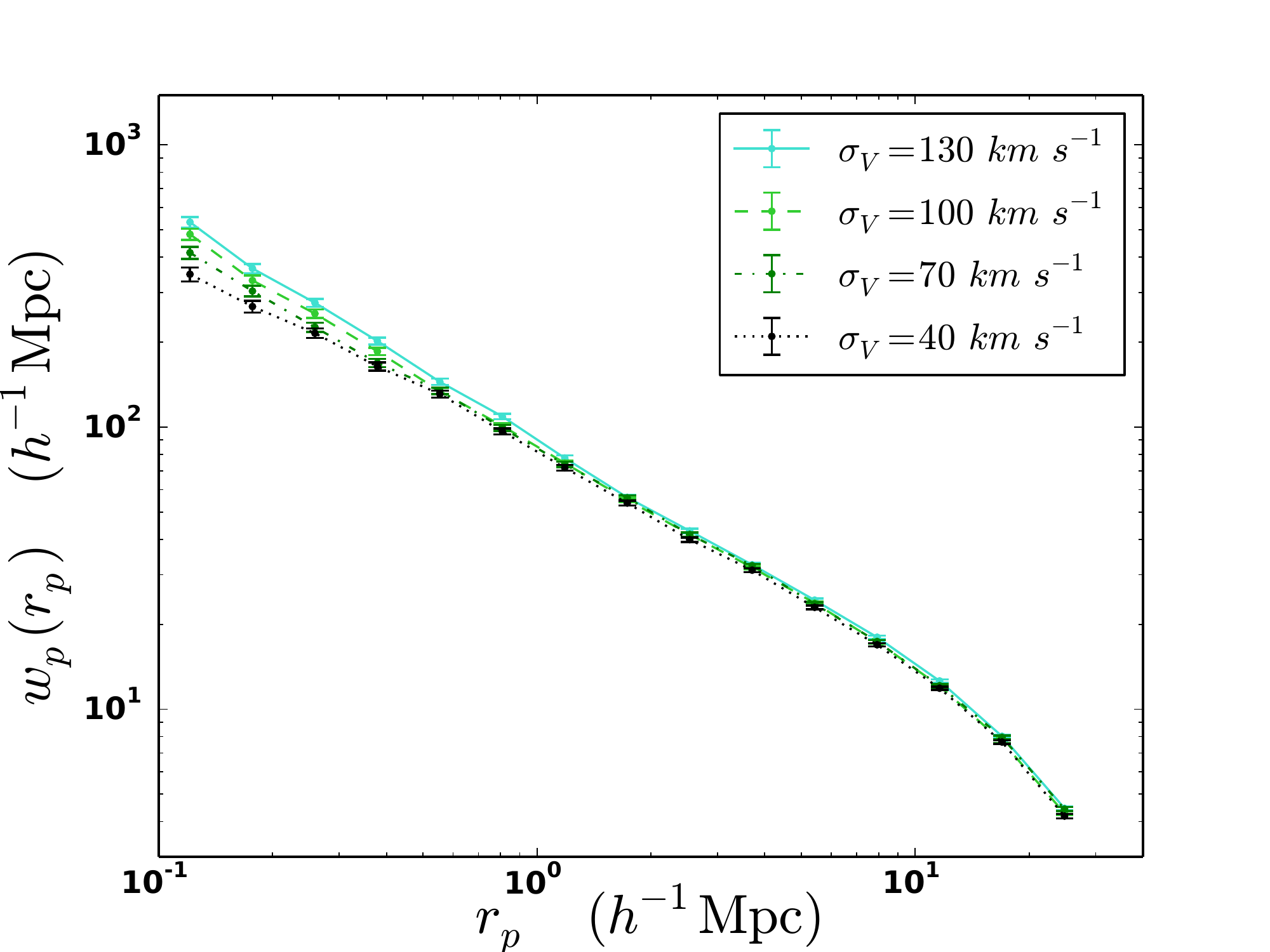}\quad
    \includegraphics[width=0.43\linewidth]{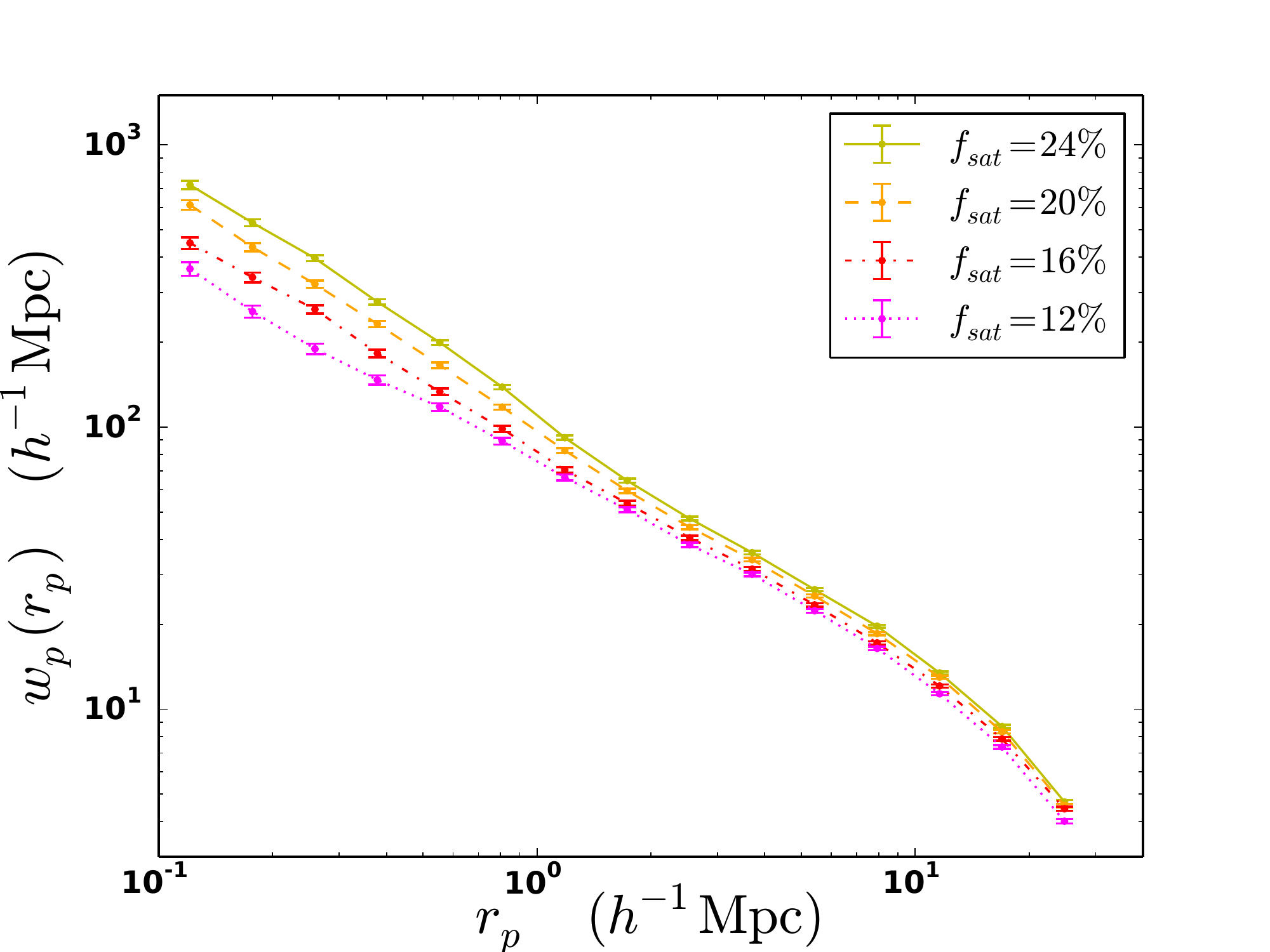}
    \caption{[OII] ELG clustering dependence on our model parameters: $V_{\rm{peak}}$ (top left), $\sigma_V$ (top right) and $f_{\rm{sat}}$ (bottom). In each panel we let vary only one parameter at a time and the other two are fixed at the best-fit values: $V_{peak}=303$\,km\,$s^{-1}$, $\sigma_V=140$\,km\,$s^{-1}$ and $f_{sat}=18\%$.}
    \label{fig:params}
\end{figure*}

\noindent Using this model, we fit the projected 2PCFs of the [OII] ELG volume-limited samples shown in Figure \ref{fig:samplesO2} and derive the optimal $V_{\rm{peak}}\pm\sigma_V$ and $f_{sat}$ values reported in Table \ref{tab:vlsO2}. The monopole and quadrupole models are then computed from the $w_p(r_p)$ best-fit mocks. For each mock catalog, we also display the most  probable (i.e. more representative) or ``typical'' $V_{\rm{peak}}$ and M$_h$ value, corresponding to the peak of the halo velocity and mass distributions. The difference between the best-fit and typical velocities is due to the fact that for certain $V_{\rm{peak}}$ values we do not have enough haloes to draw from the Gaussian PDF, then the algorithm picks haloes with smaller velocity until the desired number density is achieved. This procedure distorts the resulting mock velocity and mass distributions making them skewed, and the effect increases for those redshift slices with higher number density. This skewness motivates the choice of the most probable values (solid vertical lines in Figure \ref{fig:distrib}) as the most representative or ``typical'' for ELGs, instead of the mean values (dashed).
\noindent The variation of the SHAM scatter parameter, $\sigma$, is accounted for in the  assignment, but its effect is highly degenerate with $V_{\rm{peak}}$ and $\sigma_V$. 
A certain degree of degeneracy is also present between $V_{\rm{peak}}$, $\sigma_V$ and $f_{\rm{sat}}$, as shown in Figure \ref{fig:params}. Here we display the variation of the projected 2PCF with a parameter at a time, while the other two quantities are fixed at their best-fit values: $V_{\rm{peak}}=303$\,km\,$s^{-1}$, $\sigma_V=140$\,km\,$s^{-1}$ and $f_{\rm{sat}}=18\%$. As expected, increasing the satellite fraction we observe an enhancement of the clustering 1-halo term (bottom panel) corresponding to scales at which the correlation between substructures of the same halo start to dominate. A similar, more moderate effect is observed if we increase $\sigma_V$ (top right plot). On the other side, a modification in $V_{\rm{peak}}$ (top left) affects the 2PCF both on small and intermediate scales: higher (lower) velocity values return a higher (lower) clustering amplitude. Overall, the major variations in the clustering amplitude are driven by the satellite fraction at $r_p\lesssim1\,h^{-1}$Mpc and by  $V_{\rm{peak}}$ at  $r_p\gtrsim1\,h^{-1}$Mpc, but our SHAM implementation shows also a clear dependence on the Gaussian width $\sigma_V$, differently to the model proposed by \citet[]{Favole2016}. Despite the model parameters show some degeneracy, our modified SHAM approach is able to clearly identify a minimum in the parameter space only by fitting the 2PCFs, without any additional independent measurement. Previous works failed in this sense:  \citet[]{Favole2016} combine weak lensing with clustering measurements to rule out most of the lower and higher mass models and break the degeneracy, while \citet[]{2017MNRAS.468..728R} cannot even find a region of minima in the parameter space. 
The procedure described above guarantees the reliability of our model galaxies, since it incorporates the scatter observed between halo velocities and galaxy luminosities (encoded in the SHAM scatter parameter, $\sigma$), and allows to correctly reproduce both the ELG number density and the clustering amplitude, as shown in Figure \ref{fig:wpELGfig}.


\section{Results}
\label{sec:results}
\subsection{Clustering versus [O\textsc{II}] ELG luminosity}
\label{sec:clust}
\begin{figure*}
\centering
    \includegraphics[width=0.43\linewidth]{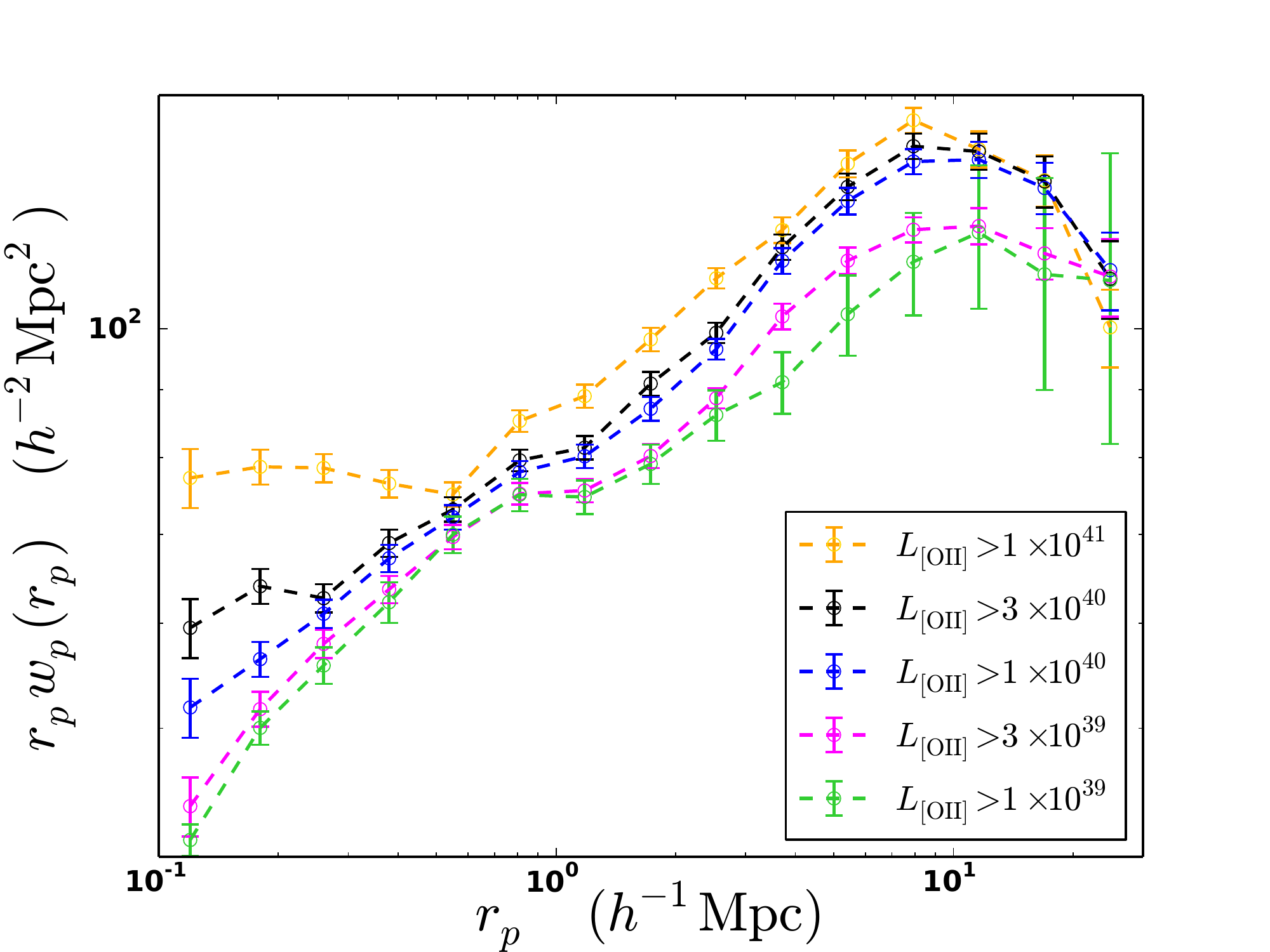}\quad
        \includegraphics[width=0.43\linewidth]{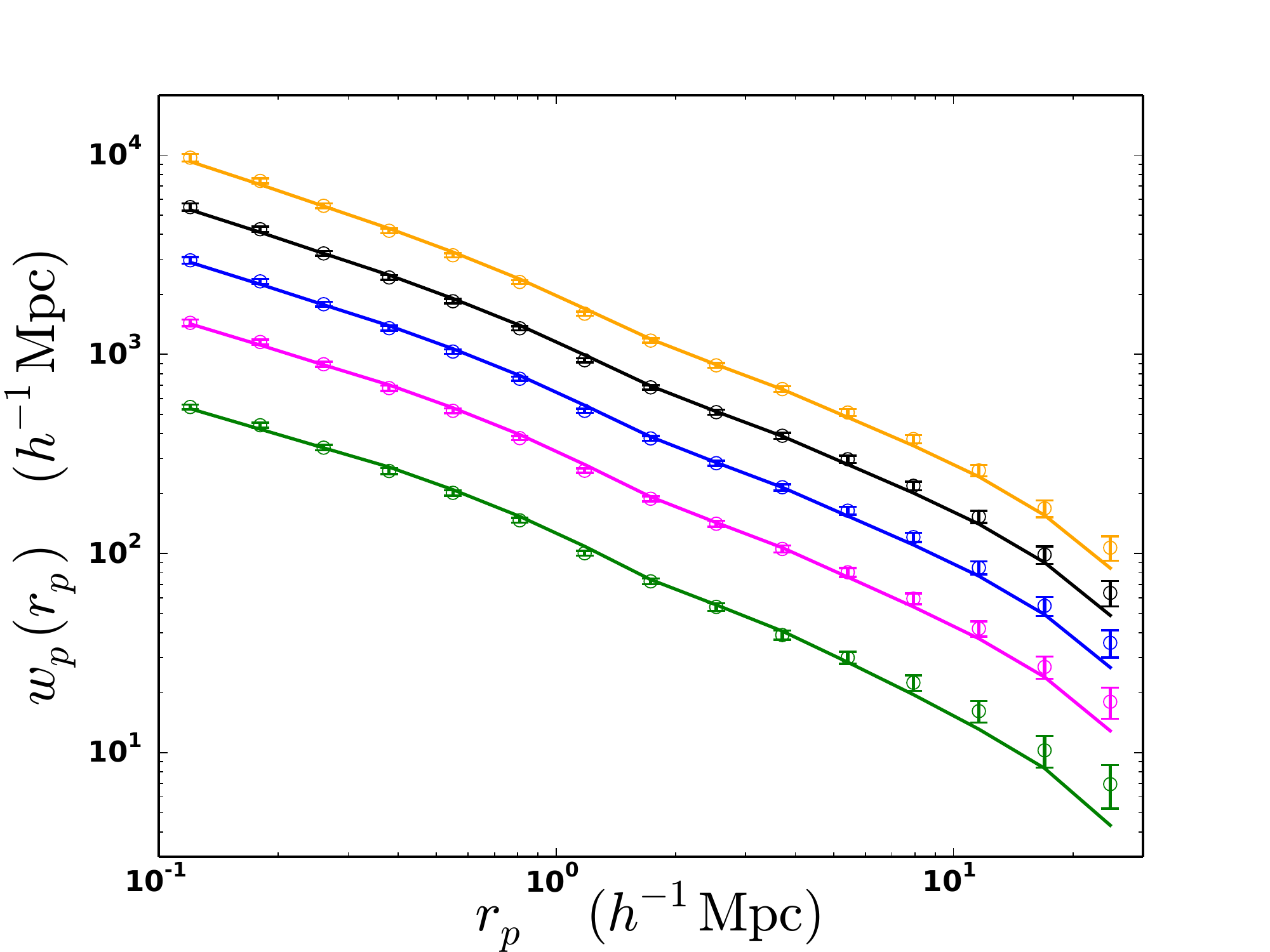}\\
    \includegraphics[width=0.43\linewidth]{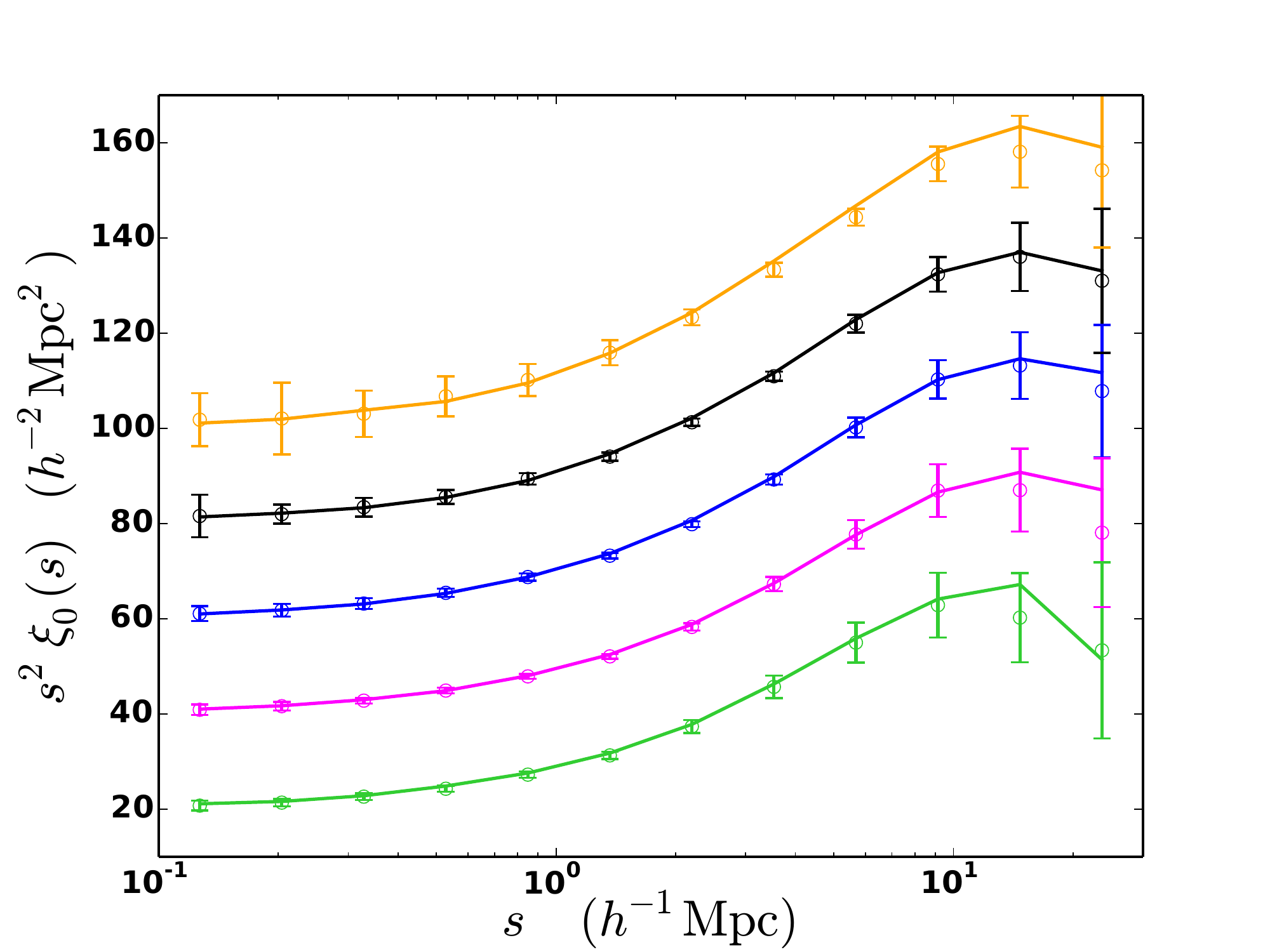}\quad
        \includegraphics[width=0.43\linewidth]{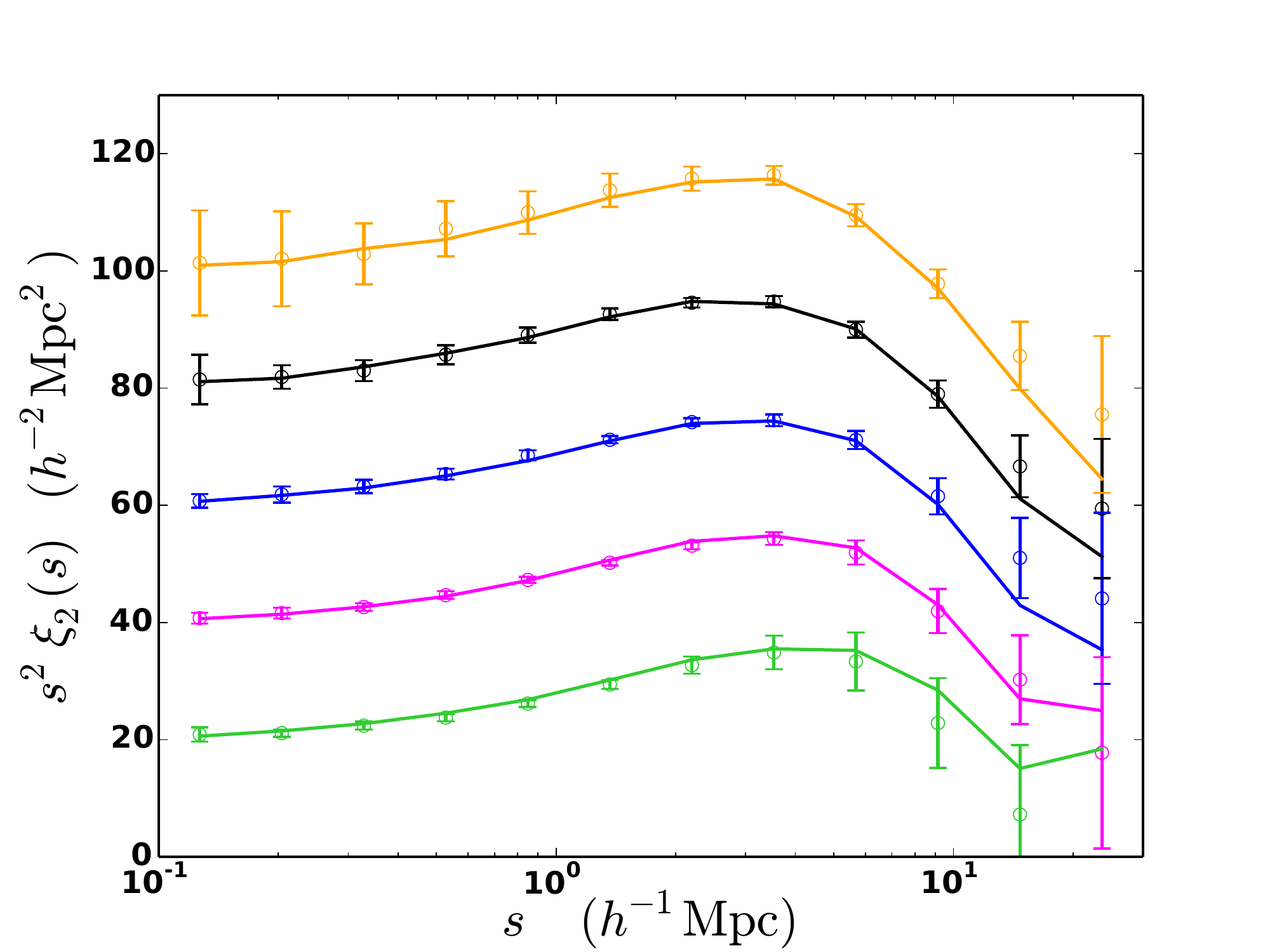}
    \caption{\textit{Top left panel:} MPA-NYU SDSS Main projected 2PCF, multiplied by the physical scale $r_p$, of the volume-limited samples in \OII luminosity thresholds defined in Table \ref{tab:vlsO2}. The MultiDark model galaxies are not shown here.  \textit{Top right:} SDSS projected 2PCFs (points)  versus our MultiDark model galaxies (lines). The errors on the measurements are estimated performing 200 jackknife re-samplings. \textit{Bottom line:} monopole (left) and quadrupole (right) correlation functions. Just for clarity, when we plot the data and the model together, we shift $w_p(r_p)$ by 0.2\,dex and $s^2\xi_{0,2}(s)$ by $20\,h^{-2}$\,Mpc$^2$ to avoid overlapping.}
    \label{fig:wpELGfig}
\end{figure*}

\noindent The MPA-NYU SDSS Main clustering measurements as a function of the \OII emission line luminosity are presented in the top left panel of Figure \ref{fig:wpELGfig}. We show the agreement with our MultiDark model galaxies in the projected (top right panel), monopole (bottom left) and quadrupole (bottom right) two-point correlation functions. When comparing data and models, we shift the $w_p(r_p)$ values by 0.2\,dex and $s^2\xi_{0,2}(s)$ by $20\,h^{-2}$\,Mpc$^2$ to avoid overlapping. We find that more luminous galaxies have a higher clustering amplitude compared to their fainter companions. Redshift and luminosity are always correlated in the construction of flux-limited samples, either radio \citep[][]{2000AJ....120...68L} or optical \citep[]{2011ApJ...736...59Z, Guo2015} or [OII] luminosity as is our case. We cannot separate the luminosity from the redshift dependence and keep having volume-limited samples. Our intention is to correctly model the galaxy clustering signal as a function of the [OII] luminosity and this dependence by construction encapsulates some evolution, which is taken into account both in the observations in the way the flux-limited samples are built, and in the models through light-cones. However, for the MPA-NYU SDSS Main galaxies the redshift evolution is relatively mild since they belong to the faint end of the [OII] luminosity function which, at $0.02<z<0.22$, is considerably flatter compared to higher z \citep[][]{2015A&A...575A..40C}.
Our SHAM predictions for the typical ELG host halo masses and satellite fractions are given in Table \ref{tab:vlsO2}, and indicate that ELGs with higher \OII luminosities tend to occupy more massive haloes, with a lower satellite fraction. We find that \OII emission line galaxies at $z\sim0.1$ live in haloes with mass between $\sim3.2\times10^{11}\,h^{-1}$M$_{\odot}$ and  $5.5\times10^{12}\,h^{-1}$M$_{\odot}$, close to to the ELG scenario found at $z\sim0.8$ by \citet[]{Favole2016}. The remarkable agreement shown in the quadrupole (Figure \ref{fig:wpELGfig}, bottom right panel) indicates that we are correctly modeling the satellite fraction. The deviations of the models from the observations beyond $10\,h^{-1}$Mpc are due to the presence of cosmic variance. 
\begin{figure}
\centering
        \includegraphics[width=0.9\linewidth]{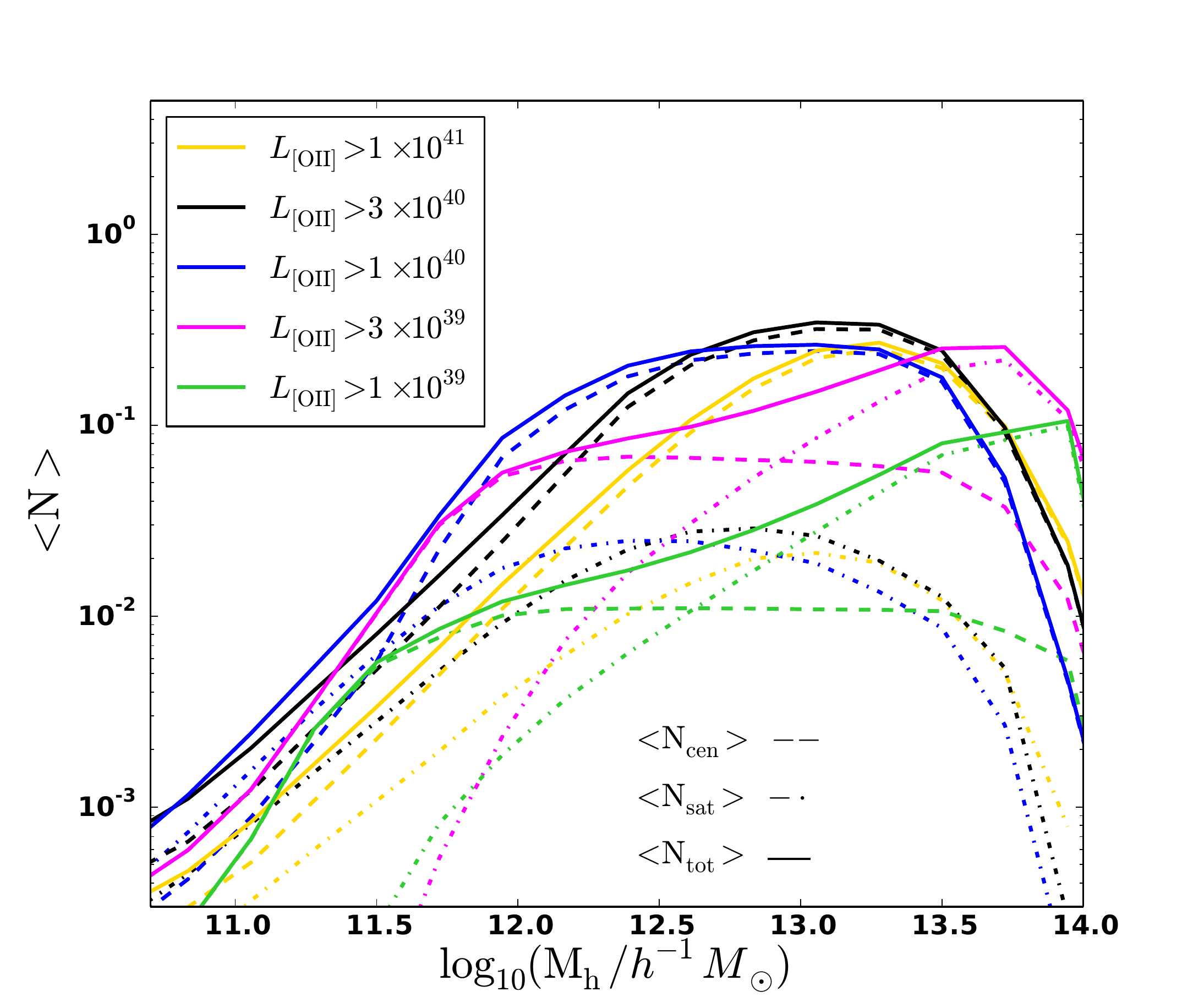}\quad
    \caption{MPA-NYU SDSS [OII] ELG halo occupation distribution, or mean number of mocks as a function of the central halo mass.}
    \label{fig:hod}
\end{figure}

\noindent The [OII] ELG halo occupation distribution obtained from the MultiDark SHAM mocks is shown in Figure \ref{fig:hod}, as a function of the parent halo mass. The mean ELG halo occupation numbers are computed by excluding those satellite galaxies whose parent is not included in the final mock catalog. This procedure prevents us to violate the fundamental HOD prescription \citep[see e.g.,][]{Cooray2002, 2011ApJ...736...59Z, Guo2015, Favole2015a}, which requires the presence of a central halo in order to have a satellite. Compared to the SDSS $r-$band magnitude scenario studied by \citet[]{Guo2015} and Favole et al. (2016), in prep., the [OII] ELG HOD functions are a factor 10 lower for the most luminous sample and a factor 100 for the dimmest one, and drop at the high-mass end (beyond $10^{13}\,h^{-1}M_{\odot}$). This is not surprising since MultiDark haloes hosting ELGs mainly belong to the low-mass domain (see Table \ref{tab:vlsO2}), thus most of the galaxies of the full light-cone lie outside the narrow mass range imposed by the ELG models. The declining shape of the HOD is consistent with several models \citep[e.g.,][]{2013A&A...557A..66B,2013ApJ...778...98S, Zehavi2005, 2011ApJ...736...59Z} for SDSS galaxies and quasars. In addition, the limited volume of the light-cone causes the sharp cut right before $\log M_h(h^{-1}M_{\odot})=14$ in the less luminous sample, which has also the smallest volume. Except for the two less luminous models, where we have selection effects due to the small size of the volume considered, the satellite HOD functions reflects the fact that more luminous emission line galaxies are hosted by more massive halos, with lower satellite fraction. However, the steep slope of their $\rm{<N_{sat}>}$ contributions towards the higher-mass end compensates the ``lack'' of satellites below $10^{13}\,h^{-1}M_{\odot}$, returning overall higher values of satellite fraction compared to the other three mocks, see Table \ref{tab:vlsO2}.

\subsection{Galaxy bias }
\label{sec:bias}

\begin{figure*}
\centering
        \includegraphics[width=0.43\linewidth]{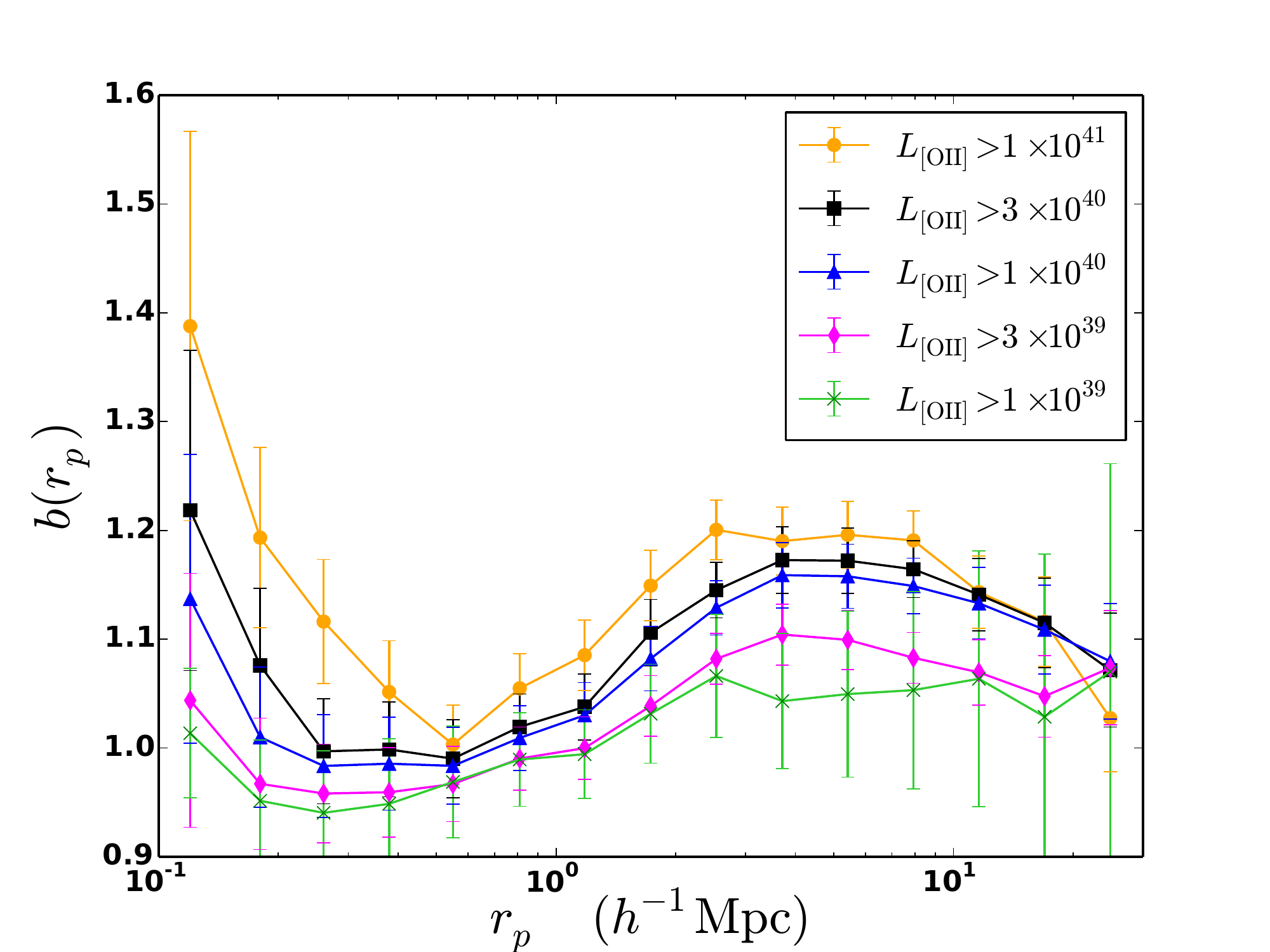}\quad
                 \includegraphics[width=0.43\linewidth]{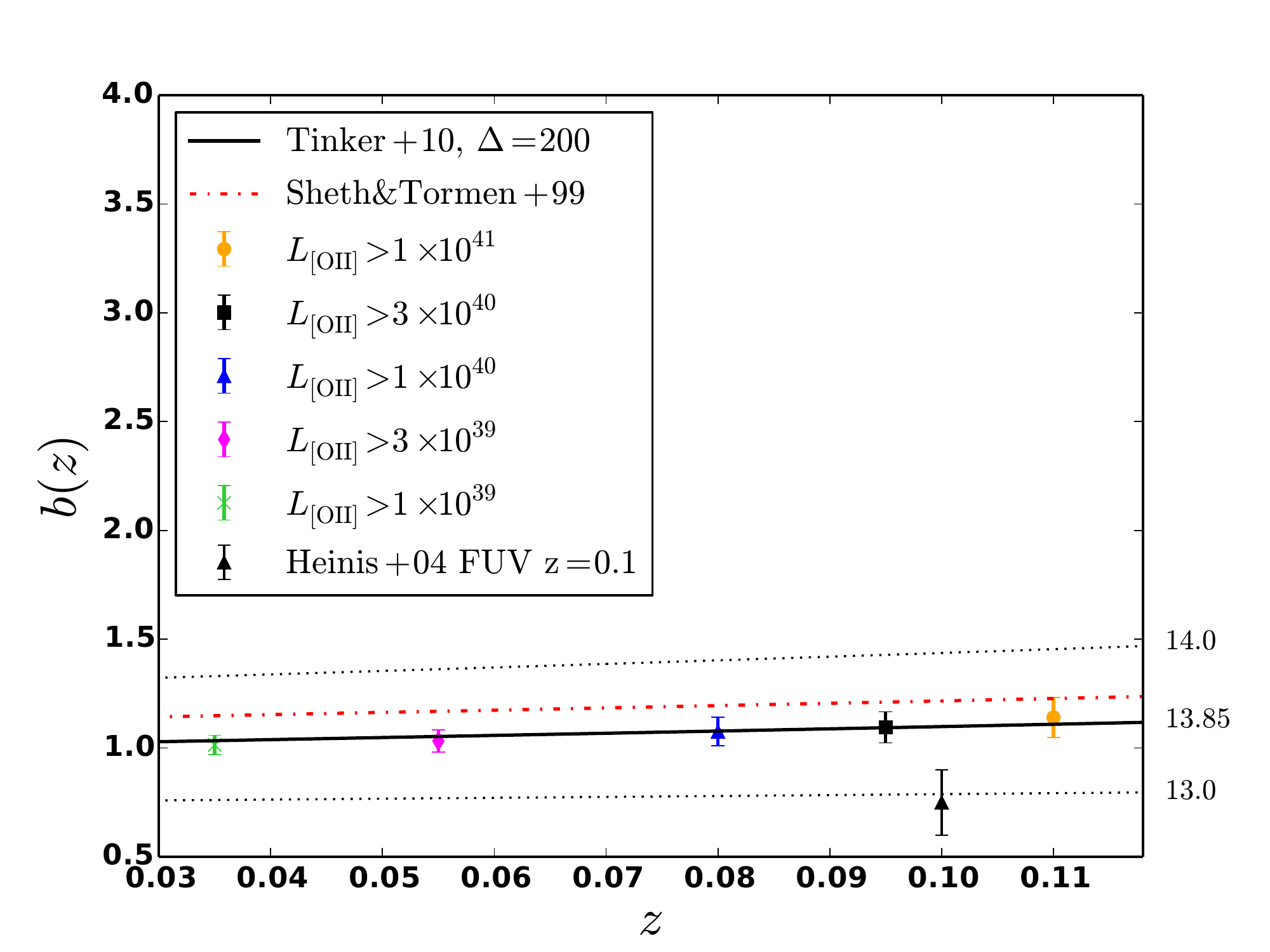}
    \caption{\textit{Left:}  Bias as a function of the physical scale for each one of the $L_{\rm{[OII]}}$ ELG volume-limited samples. \textit{Right:} MPA-NYU SDSS mean galaxy bias as a function of the mean redshift for the five volume-limited samples given in Table \ref{tab:vlsO2}. In the local Universe the bias increases with both [OII] luminosity and redshift. The dotted lines are the \citet[]{2010ApJ...724..878T} predictions assuming $\log M_{200}/h^{-1}M_{\odot}=13,14$ respectively. }
    \label{fig:bias}
\end{figure*}

\begin{figure}
\centering
         \includegraphics[width=0.9\linewidth]{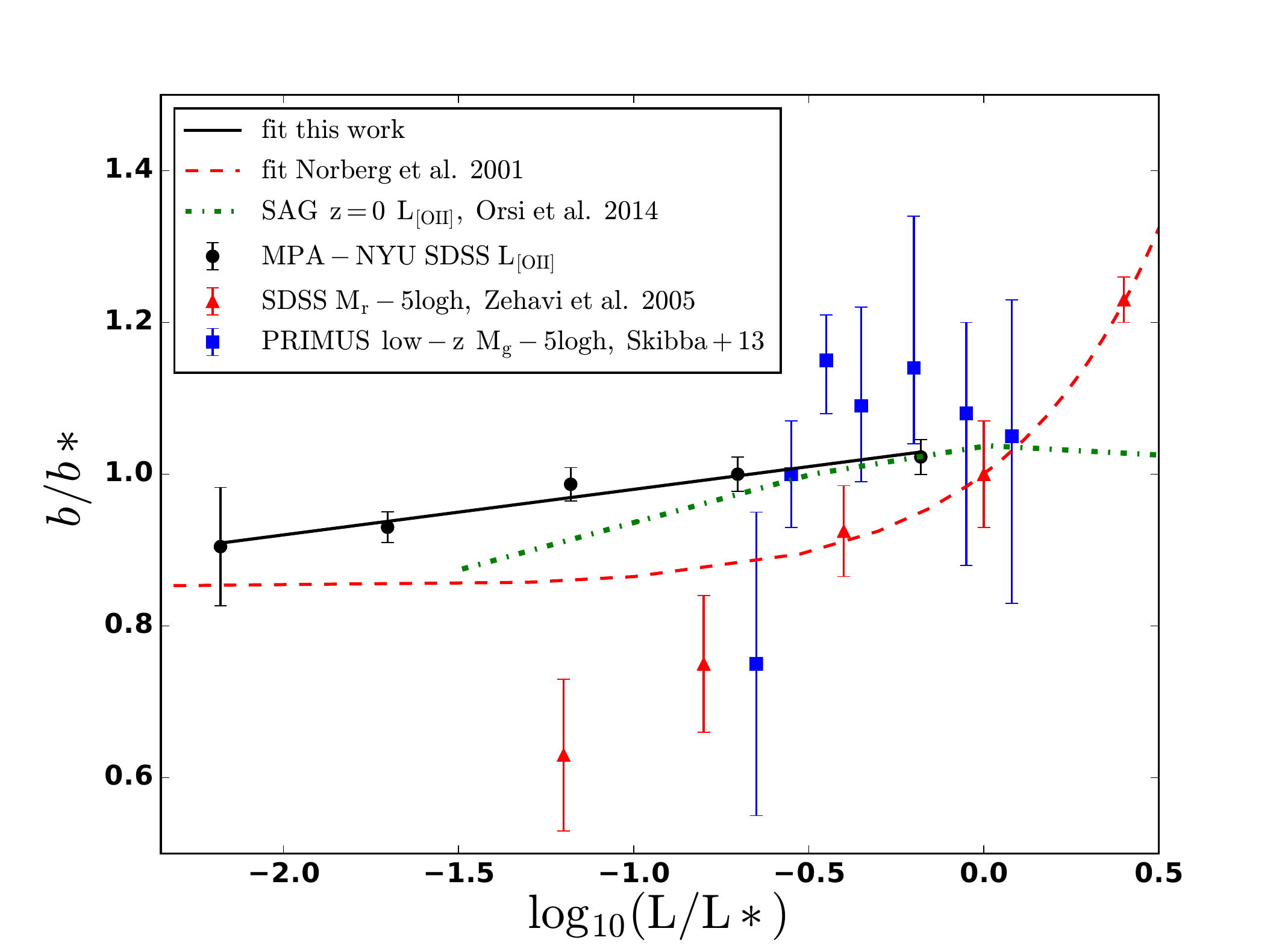}
    \caption{Normalized linear bias as a function of the SDSS [OII] emission line galaxies (black point) versus linear fit (black solid line), compared \citet[]{Zehavi2005} result for SDSS galaxies at $z\sim0.1$ (red triangles) as a function of the $r$-band luminosity. The red dashed line is the \citet[]{2001MNRAS.328...64N} fit. Here $b*$ is the bias of the ELG sample with luminosity compatible with the characteristic Schechter value $L*$ of the current \OII luminosity function \citep[see][]{2015A&A...575A..40C}. The blue squares are the PRIMUS $0.2<z<1$ measurements by \citet[]{2014ApJ...784..128S} as a function of the $g$-band magnitude. The green dot-dashed line is the prediction from \OII SAG model galaxies at $z=0$ by \citet[]{2014MNRAS.443..799O}.}
    \label{fig:biasred}
\end{figure}

\noindent To quantify the discrepancy between the observed ELG clustering signal and the underlying dark matter distribution, we compute the galaxy bias as a function of the physical scale as \citep[e.g.,][]{Nuza2013}
\begin{equation}
b(r_p)=\sqrt{w_p(r_p)/w_p^m(r_p)},
\end{equation}
where $w_p(r_p)$ is the projected 2PCF for each one of the ELG samples given in Table \ref{tab:vlsO2}, and $w_p^m(r_p)$ is the matter correlation function computed from the MDPL particle catalogue with redshift closer to $z=0.1$. The result is shown in the left panel of Figure \ref{fig:bias}, indicating that galaxy bias and \OII emission line luminosity are well correlated in the local Universe: more luminous ELGs are more biased than fainter ones. The same conclusion comes out from the right panel, where we display, for each \OII sample represented by a single point, the mean galaxy bias versus mean redshift. The theoretical curves are the predictions from \citet[]{2010ApJ...724..878T} (solid line) and \citet[]{1999MNRAS.308..119S} (dot-dashed). Assuming a mean density of $\Delta=200$ times the background and a halo mass of $\log (M_{200}/h^{-1}M_{\odot})=13.85$, the first model fits remarkably well the SDSS \OII ELG bias at $z<0.22$. However, this value is well above the ``typical'' ELG host halo masses obtained from our SHAM analysis, $\log (M_h/h^{-1}M_{\odot})\sim11.5-12.7$, indicating that the ``typical'' halo mass does not represent well the bias, which is driven by the mean mass of the halo (parent halo) for centrals (satellites).  
The dotted lines are the \citet[]{2010ApJ...724..878T} predictions assuming $\log (M_{200}/h^{-1}M_{\odot})=13,14$ respectively. The black triangle, corresponding to \citet[]{Heinis2007} SDSS-GALEX bias measurement at $z=0.1$ as a function of the rest-frame FUV luminosity, lies below our SDSS measurements due to selection effects. In fact, we select \OII emission line galaxies on top of a SDSS selection which is quite bright, while \citet[]{Heinis2007} select ELGs in the UV by imposing color and magnitude cuts to isolate bright emission line galaxies with low dust. These cuts sample the \OII luminosity function more completely than we do and typically return small haloes with low mass and low bias \citep[see e.g.,][]{Milliard2007,2013MNRAS.433.1146C, Favole2016}. In other words, thanks to our selection criterion, we are missing all the \OII emitters with low bias.

\noindent In Figure \ref{fig:biasred}, we show the normalized linear bias of the MPA-NYU SDSS emitters (black points) as a function of the [OII] luminosity. Each point corresponds to the bias of one of the ELG samples at fixed separation of $r_p=8\,h^{-1}$Mpc, that is out of the extremely nonlinear regime and where all the 2PCFs are well measured. Such a quantity is then normalized by the bias $b*$ (at the same $r_p$ separation) of the ELG sample with luminosity compatible with the characteristic Schechter value, $\log_{10}(L*/{\rm{erg\,s^{-1}}})=41.18$, of the current [OII] luminosity function at $z\sim0.1$ \citep{2015A&A...575A..40C}. All our three brightest ELG populations -- blue, black and yellow points in Figure \ref{fig:bias} -- are consistent with $L*$; we then choose the sample with $L_{\rm{[OII]}}>3\times10^{40}$erg s$^{-1}$ (black) to normalize the bias of the others.
The correlation between linear bias and \OII ELG luminosity can be easily represented by a straight line (black solid line in the right plot) $b/b*=a\log_{10}{(L_{\rm{[OII]}})}+b$, with $a=0.06\pm0.02$ and $b=1.04\pm0.02$ ($\chi^2=0.87$). Comparing our [OII] ELG bias results to the SDSS measurements by \citet[]{Zehavi2005} as a function of the $r$-band magnitude (red triangles in the plot), well fitted by \citet[]{2001MNRAS.328...64N} (red dashed line), we deduce that galaxy bias correlates more strongly with SDSS $r$-band than [OII] luminosity in the local Universe. In the same plot we show the PRIMUS bias measurement at $0.2<z<1$ \citep[]{2014ApJ...784..128S} as a function of the $g$-band absolute magnitude (blue squares). The green dot-dashed line is the [OII] ELG prediction from SAG model galaxies at $z=0$ \citep[]{2014MNRAS.443..799O}. This is compatible with our SDSS MPA-NYU results for luminosities in the range $L*/4\lesssim L_{\rm{[OII]}}\lesssim L*$.

\section{Discussion and conclusions}
\label{sec:disc}

We have presented a straightforward method to produce reliable clustering models, completely based on the MultiDark simulation products, which allows to characterize the clustering properties of the SDSS DR7 \OII emission line galaxies in $0.02<z<0.22$, by constraining their host halo masses and satellite fraction values. Our model takes the MDPL snapshots available in the redshift range of interest and generates a light-cone using the  \textsc{SUGAR} \citep[]{Rodriguez-Torres2016} algorithm. The advantage of building a light-cone is that it includes the redshift evolution and those volume effects, as the cosmic variance and the galaxy number density fluctuations, that a single simulation snapshot cannot capture. To place the observed galaxies in the MDPL halos, we apply a modified \citep[]{2017MNRAS.468..728R} SHAM prescription that accounts for the ELG stellar mass incompleteness, in which mocks are drawn through two separate PDFs (given in Eq. \ref{eq:pdfs}), one for central and one for satellite galaxies, based on a Gaussian realization. This latter is defined in terms of two parameters: the halo maximum circular velocity $V_{\rm{peak}}$ (the Gaussian mean), $\sigma_V$ (the standard deviation), and normalized to reach the observed galaxy number density, varying also the satellite fraction.  
Since there are velocity values at which we do not have enough haloes to fill the Gaussian distributions, the algorithm keeps picking haloes with lower velocities until the number density requirement is fulfilled. This condition provokes a distortion in the PDFs, and this effect is stronger for redshift bins with higher mock number density. Because of this skewness, we identify the peak (and not the mean) of each distribution as the most probable or ``typical''  host halo $V_{\rm{peak}}$ and mass value for our ELG samples. The corresponding $\sigma_V$ is estimated by computing the velocity interval around the velocity peak  in which fall 68\% of the mocks. To optimize the small-scale clustering fit, the fraction of satellite mocks is also let free to vary. As shown in Figure \ref{fig:params}, the model parameters ($V_{\rm{peak}}$, $\sigma_V$, $f_{\rm{sat}}$) suffer some degeneracy and the clustering is dominated by the halo velocity and satellite fraction. However, differently to the SHAM implementation proposed by \citet[]{Favole2016}, our approach reveals a clear dependence on the Gaussian width $\sigma_V$ and allows us to fix its value by fitting only the 2PCFs, without any additional weak lensing constraint. 

\noindent Our SDSS galaxy data and MultiDark mock catalogues are publicly available for the community on the \textsc{Skies and Universes} database\footnote{\url{http://projects.ift.uam-csic.es/skies-universes/SUwebsite/indexSDSS_OII_mock.html}}, and also as \textsc{MNRAS} online tables.

\noindent The current analysis reveals that emission line galaxies at $z\sim0.1$ with stronger \OII luminosities have higher clustering amplitudes and live in more massive haloes with lower satellite fraction. We find that the ELG bias correlates with both \OII luminosity and redshift: the stronger the \OII luminosity, the more biased the galaxy, the higher the redshift. Compared to previous studies \citep[e.g.,][]{Zehavi2005, 2014ApJ...784..128S} of the correlation between galaxy bias and luminosity,  we find that the ELG bias at $z\sim0.1$ correlates less steeply with the \OII emission line luminosity than the SDSS $r$-band or $g$-band absolute magnitudes (see Figure \ref{fig:biasred}). 

\noindent In the future, we plan to explore the dependence of the ELG clustering on the star formation rate and its evolution with redshift using the MultiDark Galaxies\footnote{\url{www.multidark.org}} new products, which will be released soon. This latter is a project, currently in development, which combines the MDPL DM-only simulation products with semi-analytic models of galaxy formation as SAG, SAGE and GALFORM (see also Favole et al. 2016, in prep.). Within the next two years, the eBOSS \citep[]{2016AJ....151...44D} survey will observe about 200,000 ELGs, which will allow us to increase the accuracy in our measurements and precisely determine the typical ELG host halo masses, satellite fraction values and velocity distributions. At the same time, the Low Redshift survey at Calar Alto \citep[LoRCA;][]{2016MNRAS.458.2940C} plans to observe galaxies at $z<0.2$ in the northern sky, to complement the current SDSS-III/BOSS and SDSS-IV/eBOSS database. These projects will provide accurate spectroscopy and photometry for a huge number of galaxy targets, allowing us to considerably improve the quality of our emission line galaxy measurements and the precision of our models.


\section*{Acknowledgments}

GF is supported by a European Space Agency (ESA) Research Fellowship at the European Space Astronomy Center (ESAC) in Madrid, Spain.

\noindent GF and CC acknowledge financial support from the Spanish MICINN Consolider-Ingenio 2010 Programme under grant MultiDark 
CSD2009 - 00064, MINECO Centro de Excelencia Severo Ochoa Programme under grant SEV-2012-0249, and MINECO grant AYA2014-60641-C2-1-P.
JC acknowledges financial support from MINECO (Spain) under project number  AYA2012 - 31101. 

\noindent The MultiDark Planck  simulation  has  been performed in the Supermuc supercomputer at
the Libniz Supercomputing Center (LRZ, Munich) thanks to the cpu time awarded by PRACE (proposal number 2012060963).

\noindent Funding for the SDSS and SDSS-II has been provided by the Alfred P. Sloan Foundation, the Participating In- stitutions, the National Science Foundation, the U.S. De- partment of Energy, the National Aeronautics and Space Administration, the Japanese Monbukagakusho, the Max Planck Society, and the Higher Education Funding Council for England. The SDSS Web Site is \url{http://www.sdss.org/}. 
The SDSS is managed by the Astrophysical Research Consortium for the Participating Institutions. The Partic- ipating Institutions are the American Museum of Natu- ral History, Astrophysical Institute Potsdam, University of Basel, University of Cambridge, Case Western Reserve University, University of Chicago, Drexel University, Fermilab, the Institute for Advanced Study, the Japan Participation Group, Johns Hopkins University, the Joint Institute for Nuclear Astrophysics, the Kavli Institute for Particle As- trophysics and Cosmology, the Korean Scientist Group, the Chinese Academy of Sciences (LAMOST), Los Alamos Na- tional Laboratory, the Max-Planck-Institute for Astronomy (MPIA), the Max-Planck-Institute for Astrophysics (MPA), New Mexico State University, Ohio State University, Uni- versity of Pittsburgh, University of Portsmouth, Princeton University, the United States Naval Observatory, and the University of Washington. 

\bibliographystyle{mn2e}
\bibliography{./references}

\end{document}